\documentclass[10pt]{article}
\usepackage[preprint]{tmlr}
\usepackage{amsmath,amssymb}
\usepackage{booktabs,tabularx}
\usepackage{graphicx}
\usepackage{flafter,placeins}
\usepackage{tikz}
\usetikzlibrary{arrows.meta,positioning,calc}
\usepackage{hyperref}
\usepackage{url}
\hypersetup{hidelinks}

\title{Harmony: Bridging Audio Understanding and Editing\\through Agentic Collaboration}\author{\name Zhongyi Tong \email ztong@wpi.edu \\
      \addr Worcester Polytechnic Institute
      \AND
      \name Yapeng Tian \email yapeng.tian@utdallas.edu \\
      \addr The University of Texas at Dallas}

\begin{document}
\maketitle

\begin{abstract}
Natural-language audio editing requires connecting a user's intent to specific sounds and time regions while preserving surrounding content. Complex requests make this connection demanding: an edit may select one occurrence, depend on a temporal relation, or modify a source that overlaps with other sounds. We introduce Harmony, an agentic framework that bridges audio understanding and editing through collaboration among an LLM planner, an audio-language model, and external tools. Harmony combines two mechanisms. Edit-driven collaborative planning uses targeted questions and tool measurements to determine the target, timing, and operation parameters. Evidence-guided execution uses observations of the first separated or generated audio to trigger bounded corrections before final mixing. Together, these mechanisms connect semantic decisions to waveform operations throughout the editing process. Harmony supports addition, removal, replacement, and volume adjustment. On 420 MMAE examples spanning these operations, Harmony achieves an exact match rate of 22.62\%, compared with 13.57\% for the strongest baseline, and improves aggregate instruction following and context preservation.
\par\smallskip
\noindent\textbf{Project website:} \url{https://christong9.github.io/harmony/}
\end{abstract}

\begingroup
\renewcommand{\thefootnote}{}
\begin{NoHyper}
\footnotetext{ChatGPT was used to assist with drafting and revising this manuscript. The authors take full responsibility for its content, including all claims, results, and references.}
\end{NoHyper}
\endgroup

\section{Introduction}
\label{sec:introduction}

Natural-language audio editing allows users to request changes to a recording without specifying the signal-processing operations needed to produce them. The goal is to realize these changes while preserving the content and temporal structure that should remain intact. Consider the request, ``Replace each dog bark with a cat meow while keeping the conversation unchanged.'' Executing it requires locating each bark, producing suitable replacement sounds, and preserving any overlapping speech. A plausible new sound is therefore only part of a successful edit: the system must also determine what to modify, where to act, and how to integrate the result with the existing recording. Meeting these requirements depends on both the capabilities of audio models and the workflow that coordinates their use.

Instruction-conditioned models have made audio modification more accessible. Diffusion-based approaches learn to transform an input recording according to a text instruction, enabling changes to sound events and musical content~\citep{wang2023audit,han2023instructme,tao2025mmedit}. Source conditioning and inversion help retain the original content, while temporal controls support localized changes~\citep{manor2024zeta,xu2024ppae}. These methods bring powerful generative capabilities to editing, but producing the requested change and preserving the rest of the recording remain distinct requirements. Reconstructing audio can introduce changes beyond the intended edit, even when the output retains its overall content and structure~\citep{manor2024zeta,ellis2025recomposer}. Consequently, complex editing instructions require precise control over both the audio content to be modified and the original content that must be preserved.

At the system level, agentic frameworks organize editing into explicit operations. WavCraft combines specialist audio models with signal-processing tools, while SmartDJ decomposes requests into operations performed by a diffusion editor~\citep{liang2024wavcraft,lan2025smartdj}. This flexibility allows a workflow to combine generation, separation, and direct manipulation of existing audio. Reliable editing, however, also depends on the evidence used to choose and carry out these operations. Selecting the second occurrence of a sound or placing an event after an utterance requires more than a general description of the recording; the system must identify the relevant occurrence and its boundaries. WavCraft itself notes the difficulty of extracting precise temporal relationships from audio~\citep{liang2024wavcraft}. Execution creates a further dependency: if separation leaves the old sound in a replacement window, adding the requested new sound does not complete the edit. The system must therefore connect instruction-specific observations to planning and account for the intermediate audio its tools actually produce.

We introduce \textbf{Harmony}, an agentic framework that connects audio understanding and editing through two complementary mechanisms. \emph{Edit-driven collaborative planning} coordinates an LLM planner, Qwen3-Omni~\citep{qwen3omni}, and external analysis tools to select the sounds to edit and determine their temporal scope. \emph{Evidence-guided execution} uses observations of the initial separation or generation outputs to select bounded corrections before placement and mixing. These corrections address mismatches in the intermediate audio while retaining the instruction's target and preservation requirements.

Our contributions are threefold:
\begin{enumerate}
    \item \textbf{Edit-driven collaborative planning.} We introduce a planning mechanism that coordinates an LLM, Qwen3-Omni, and external analysis tools to connect editing instructions to selected sources, time windows, and operation parameters.
    \item \textbf{Evidence-guided execution.} We develop an execution mechanism that uses intermediate audio observations to select bounded corrective actions on separated or generated material, retaining the planned target and preservation requirements.
    \item \textbf{Evaluation across four editing operations.} On all 420 eligible MMAE examples~\citep{ma2026mmae}, covering addition, removal, replacement, and volume adjustment, Harmony leads the evaluated baselines in overall IFR, CR, and EMR and achieves the highest EMR for each operation. A $2\times2$ ablation examines the roles of external evidence acquisition and conditional correction.
\end{enumerate}

\section{Related Work}
\label{sec:related-work}

\paragraph{Instruction-conditioned audio editing.}
Audio editing combines a requested change with preservation of relevant content in an existing recording. AUDIT learns instruction-guided transformations with a latent diffusion model, while InstructME studies music editing and remixing with attention to musical consistency \citep{wang2023audit,han2023instructme}. SAO-Instruct adapts Stable Audio Open to free-form editing instructions using paired training examples \citep{ungersbock2025saoinstruct}. MMEdit jointly represents audio and instructions with an audio language model to condition a diffusion editor \citep{tao2025mmedit}. Beyond diffusion models, Instruct-MusicGen adapts MusicGen for adding, removing, and extracting musical stems \citep{zhang2025instructmusicgen}. Training-free methods offer another route: ZETA applies DDPM inversion to pretrained audio diffusion models, and PPAE manipulates cross-attention maps for local edits \citep{manor2024zeta,xu2024ppae}. More recent work explicitly addresses event timing. Recomposer combines textual edit descriptions with an event-roll representation to control the action, sound class, and temporal extent \citep{ellis2025recomposer}. FreeSonic extracts target regions from text--audio attention and confines modifications through attention decoupling \citep{jiang2026freesonic}. 

\paragraph{Agentic frameworks for audio editing.}
WavCraft combines specialist audio models and signal-processing operations into executable programs, while SmartDJ decomposes high-level requests into atomic edits performed by a diffusion editor~\citep{liang2024wavcraft,lan2025smartdj}. RIME supports iterative music post-production through the POEMS toolkit and an agent that listens to intermediate audio and updates its plan~\citep{schaffer2026rime}. AVE-Agent coordinates audio--visual editing through subtask evaluation, prompt revision, and feedback-guided replanning~\citep{wen2026avecompass}. Harmony binds selected events and tool-derived timestamps to an executable edit plan, then applies bounded corrections to separated or generated audio before mixing. These corrections retain the selected target and execution windows.

\begin{figure}[t]
\centering
\definecolor{harmonyInk}{HTML}{263A50}
\definecolor{harmonyTeal}{HTML}{328D87}
\definecolor{harmonyAmber}{HTML}{CB963D}
\definecolor{harmonyViolet}{HTML}{7772AF}
\begin{tikzpicture}[
    font=\sffamily\fontsize{9}{10.5}\selectfont,
    box/.style={draw=harmonyInk!70,line width=.65pt,rounded corners=2pt,
        fill=white,align=center,inner sep=3pt,minimum height=8mm},
    flow/.style={-{Latex[length=2.2mm,width=1.5mm]},line width=1pt,draw=harmonyInk},
    opt/.style={flow,dashed},
    lab/.style={font=\sffamily\fontsize{8}{9.3}\selectfont,inner sep=1pt,text=harmonyInk},
    stage/.style={anchor=west,font=\sffamily\fontsize{10}{11.5}\selectfont\bfseries,text=harmonyInk}
]
\draw[rounded corners=4pt,draw=harmonyInk!15,fill=harmonyInk!3]
    (-7.55,4.0) rectangle (7.55,7.55);
\draw[rounded corners=4pt,draw=harmonyTeal!20,fill=harmonyTeal!3]
    (-7.55,0) rectangle (7.55,3.85);
\node[circle,fill=harmonyInk,text=white,inner sep=2.4pt,font=\sffamily\bfseries] at (-7.08,7.25) {1};
\node[stage] at (-6.7,7.25) {Edit-driven collaborative planning};

\node[box,text width=29mm,minimum height=6mm] (xlisten) at (-5.0,6.6) {};
\draw[harmonyTeal,line width=.9pt] plot coordinates
    {(-6.15,6.6) (-6.06,6.6) (-6.0,6.75) (-5.92,6.44)
     (-5.84,6.82) (-5.76,6.5) (-5.68,6.68) (-5.61,6.6)};
\node at (-4.60,6.6) {Recording $x$};
\node[box,text width=33mm,minimum height=6mm] (instruction) at (0,6.6) {Instruction $u$};
\node[box,draw=harmonyTeal,fill=harmonyTeal!8,text width=29mm] (observer) at (-5.0,5.55)
    {\textbf{Qwen}\\audio observations};
\node[box,draw=harmonyInk,fill=harmonyInk!7,text width=33mm] (planner) at (0,5.55)
    {\textbf{LLM planner}\\next question / tool};
\node[box,draw=harmonyAmber,fill=harmonyAmber!10,text width=30mm] (analysis) at (5.0,5.55)
    {\textbf{Analysis tools}\\activity / ASR\\source timing};
\node[box,draw=harmonyInk,text width=92mm] (plan) at (0,4.45)
    {\textbf{Edit plan $\pi$}: operation, target occurrences, windows $W$\\parameters + preservation requirements};
\draw[flow] (xlisten) -- (observer);
\draw[flow] (instruction) -- (planner);
\draw[flow] ([yshift=1.8mm]planner.west) --
    node[lab,above=2pt]{questions} ([yshift=1.8mm]observer.east);
\draw[flow] ([yshift=-1.8mm]observer.east) --
    node[lab,below=2pt]{answers} ([yshift=-1.8mm]planner.west);
\draw[flow] ([yshift=1.8mm]planner.east) --
    node[lab,above=2pt]{tool calls} ([yshift=1.8mm]analysis.west);
\draw[flow] ([yshift=-1.8mm]analysis.west) --
    node[lab,below=2pt]{results} ([yshift=-1.8mm]planner.east);
\draw[flow] (planner) -- (plan);

\node[circle,fill=harmonyTeal,text=white,inner sep=2.4pt,font=\sffamily\bfseries] at (-7.08,3.58) {2};
\node[stage] at (-6.7,3.58) {Evidence-guided execution};
\node[box,draw=harmonyAmber,fill=harmonyAmber!10,text width=33mm] (tools) at (-5.2,2.4)
    {\textbf{Separate / generate}\\SAM Audio\\Stable Audio 3};
\node[box,text width=33mm] (material) at (-1.0,2.4)
    {\textbf{Intermediate audio}\\residual / new asset};
\node[box,draw=harmonyViolet,fill=harmonyViolet!10,text width=33mm] (mixer) at (3.5,2.4)
    {\textbf{Temporal placement}\\+ mixing};
\node[box,draw=harmonyViolet,fill=harmonyViolet!10,minimum width=9mm] (output) at (6.4,2.4) {$y$};
\node[box,dashed,draw=harmonyTeal,fill=harmonyTeal!8,text width=43mm] (correct) at (-1.0,.8)
    {\textbf{Conditional correction}\\first evidence + $u,\pi$\\fixed call budget};
\node[box,text width=27mm] (xsource) at (-5.2,.65) {Original recording $x$};
\node[box,text width=27mm] (xmix) at (4.4,.65) {Original recording $x$};

\draw[flow] ([xshift=-16mm]plan.south) -- (-1.6,3.12) -- (-5.2,3.12) -- (tools.north);
\draw[flow] ([xshift=35mm]plan.south) --
    node[lab,right=3pt]{windows + preservation} (mixer.north);
\draw[flow] (tools) -- (material);
\draw[flow] (material) -- (mixer);
\draw[flow] (mixer) -- (output);
\draw[opt] (material.south) -- node[lab,right=3pt]{if applicable} (correct.north);
\draw[opt] (correct.east) -- (2.2,.8) -- ([xshift=-13mm]mixer.south);
\draw[flow] (xsource.north) -- node[lab,left=3pt]{source audio} (tools.south);
\draw[flow] (xmix.north) -- node[lab,right=3pt]{original base} ([xshift=9mm]mixer.south);

\draw[flow] (-6.8,-.3) -- (-5.95,-.3);
\node[lab,anchor=west] at (-5.8,-.3) {data / plan flow};
\draw[opt] (-2.7,-.3) -- (-1.85,-.3);
\node[lab,anchor=west] at (-1.7,-.3) {optional correction};
\end{tikzpicture}
\caption{Harmony's two mechanisms. Collaborative planning resolves the target, timing, and preservation requirements through exchanges with Qwen and analysis tools. Execution conditions bounded corrections on the first intermediate audio and the edit plan, then places and mixes the resulting material with the original recording. Dashed paths are optional.}
\label{fig:overview}
\end{figure}
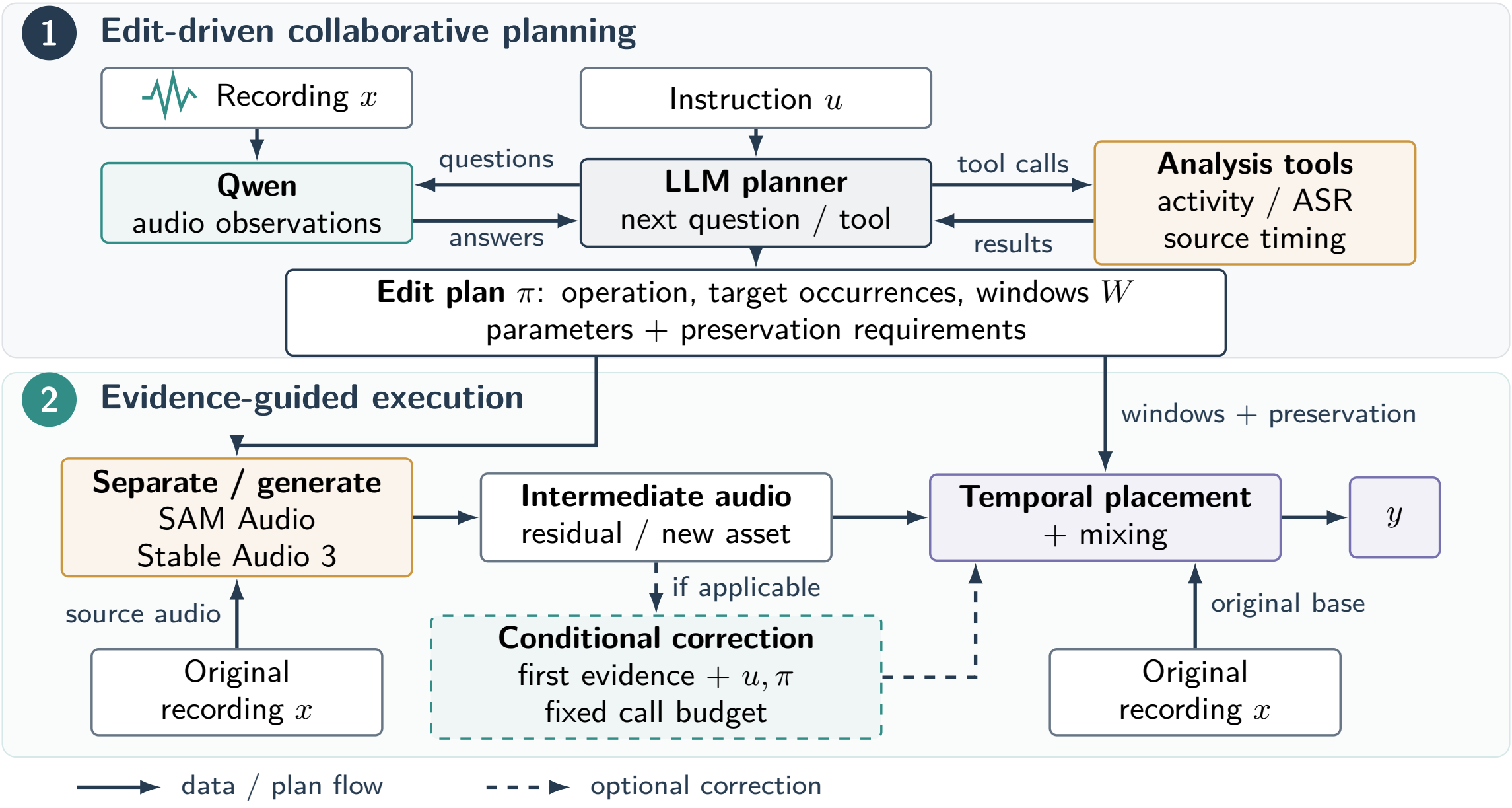

\paragraph{Tool-assisted audio understanding.}
ReAct studies the interleaving of reasoning and actions in general language agents \citep{react}. Within audio understanding, AudioToolAgent uses a central LLM to query audio-language models and speech tools, ask follow-up questions, and reconcile their outputs \citep{wijngaard2025audiotoolagent}. Audio-Mind uses a planner to request external evidence and targeted re-listening when the audio frontend reports unresolved questions \citep{wang2026audiomind}. In its reported hot-versus-cold water example, the system isolates two pouring segments, compares spectral measurements, and asks the frontend to listen again before answering. These audio-understanding systems use tools for acoustic reasoning and temporal analysis, with evaluation focused on answering questions about audio.

 \section{Methodology}
\label{sec:method}

\subsection{Task and framework overview}
\label{sec:overview}

Given a recording $x$ and a natural-language instruction $u$, Harmony produces an edited recording $y$ by determining what to change, where to act, and what to preserve. A request to replace the second bark while retaining a conversation, for example, requires selecting an occurrence, locating its boundaries, and modifying the target within overlapping audio.

Harmony organizes this process through two mechanisms (Figure~\ref{fig:overview}). \emph{Edit-driven collaborative planning} coordinates an LLM planner, Qwen3-Omni as an audio observer~\citep{qwen3omni}, and analysis tools to produce an edit plan $\pi$. The plan specifies the operation, target occurrences, temporal scope, parameters, and preservation requirements. \emph{Evidence-guided execution} uses observations of intermediate audio to select bounded corrections when separation or generation does not provide suitable material for the planned edit.

SAM Audio~\citep{samaudio} supplies separated sources and residual audio, while Stable Audio 3~\citep{stableaudio3} generates new material for addition or replacement. The resulting layers are placed and mixed according to the plan to form the final recording.

\subsection{Edit-driven collaborative planning}
\label{sec:collaboration}

The planner extracts explicit editing requirements and asks Qwen to resolve missing information about the target sounds and their timing. Qwen's responses guide the planner's follow-up questions and analysis-tool calls.

The analysis tools supply complementary temporal evidence. Waveform-energy analysis locates candidate activity intervals; separation-assisted analysis isolates a candidate source and measures its active regions; speech analysis provides transcription and word-level timing. Qwen relates these measurements to the requested sounds. For example, activity intervals may suggest several boundaries, while linguistic context determines which one marks the end of an utterance. The planner combines this semantic interpretation with measured timing to resolve event selection and temporal relations.

For tool-derived timing, the program resolves selected region or word identifiers to their stored timestamps. The planner combines these timestamps with the instruction's temporal constraints to define the execution windows $W$. The final plan specifies the target sounds, execution windows, operation parameters, and content to preserve.

If the initial plan leaves essential information unresolved, Harmony permits bounded follow-up planning while retaining the instruction's explicit constraints.

\begin{table}[!htbp]
\centering
\setlength{\tabcolsep}{4pt}
\renewcommand{\arraystretch}{1.12}
\caption{Audio evidence and corresponding execution corrections.}
\label{tab:correction_scope}
\begin{tabularx}{\textwidth}{@{}>{\raggedright\arraybackslash}p{0.20\textwidth}>{\raggedright\arraybackslash}X>{\raggedright\arraybackslash}p{0.29\textwidth}@{}}
\toprule
Correction & Evidence for selection & Action \\
\midrule
Separation-query correction & Observations of the original audio and residual support a more specific description of the same requested target. & Repeat separation using the more specific target description. \\
Retained-source extraction & The instruction removes a complete background category, and the source inventory identifies exactly one complementary source to preserve. & Extract the retained source directly. \\
Residual clearing & Initial observations identify only content designated for replacement, with no protected or uncertain sources. & Silence the residual only within the selected replacement windows. \\
Generated-asset cleanup & Observation of the generated asset identifies an unwanted old instrument or voice. & Remove the identified content from the generated asset before placement. \\
Timbre correction & A generated instrumental asset has an incompatible timbre and satisfies the conditions for single-source pitched rendering. & Transcribe and re-render the generated notes with a supported target timbre. \\
\bottomrule
\end{tabularx}
\end{table}

\begin{figure}[t]
\centering
\definecolor{harmonyTeal}{HTML}{328D87}
\definecolor{harmonyAmber}{HTML}{CB963D}
\definecolor{harmonyViolet}{HTML}{7772AF}
\begin{tikzpicture}[
    font=\sffamily\fontsize{9}{10.5}\selectfont,
    lab/.style={anchor=east,align=right},
    band/.style={draw=black!60,line width=.5pt},
    dim/.style={{Latex[length=1.5mm]}-{Latex[length=1.5mm]},line width=.6pt},
    note/.style={font=\sffamily\fontsize{8}{9.3}\selectfont}
]
\node[anchor=west,font=\sffamily\fontsize{9}{10.5}\selectfont\bfseries] at (0,3.95)
    {Schematic: replace the second bark; preserve the conversation};
\fill[harmonyViolet!6] (8.05,.05) rectangle (10.45,3.15);
\draw[dashed,black!55] (8.05,.05) -- (8.05,3.15);
\draw[dashed,black!55] (10.45,.05) -- (10.45,3.15);
\draw[dim] (2.8,3.15) -- node[note,above=2pt]{unchanged $x$} (8.05,3.15);
\draw[dim] (8.05,3.15) -- node[note,above=2pt]{selected window $W$} (10.45,3.15);
\draw[dim] (10.45,3.15) -- node[note,above=2pt]{unchanged $x$} (14.3,3.15);

\node[lab] at (2.55,2.55) {Input $x$};
\draw[band,fill=harmonyTeal!18] (2.8,2.23) rectangle (14.3,2.53);
\node[note] at (6.5,2.38) {conversation};
\draw[band,fill=harmonyAmber!30] (4.0,2.55) rectangle (5.4,2.86);
\node[note] at (4.7,2.705) {bark 1};
\draw[band,fill=harmonyAmber!30] (8.05,2.55) rectangle (10.45,2.86);
\node[note] at (9.25,2.705) {bark 2};

\node[lab] at (2.55,1.75) {Residual in $W$};
\draw[band,fill=harmonyTeal!18] (8.05,1.6) rectangle (10.45,1.9);
\node[note,anchor=west] at (10.65,1.75) {conversation to preserve};
\node[lab] at (2.55,1.15) {New material in $W$};
\draw[band,fill=harmonyViolet!25] (8.05,1.0) rectangle (10.45,1.3);
\node[note] at (9.25,1.15) {new sound};

\node[lab] at (2.55,.4) {Output $y$};
\draw[band,fill=harmonyTeal!18] (2.8,.08) rectangle (14.3,.38);
\draw[band,fill=harmonyAmber!30] (4.0,.4) rectangle (5.4,.71);
\node[note] at (4.7,.555) {bark 1};
\draw[band,fill=harmonyViolet!25] (8.05,.4) rectangle (10.45,.71);
\node[note] at (9.25,.555) {new sound};
\end{tikzpicture}
\caption{Schematic of local replacement. Removal and insertion share the selected window $W$; the residual and new material are combined within $W$, while audio outside $W$ is retained. Green bands indicate the conversation-preservation requirement.}
\label{fig:mixing}
\end{figure}
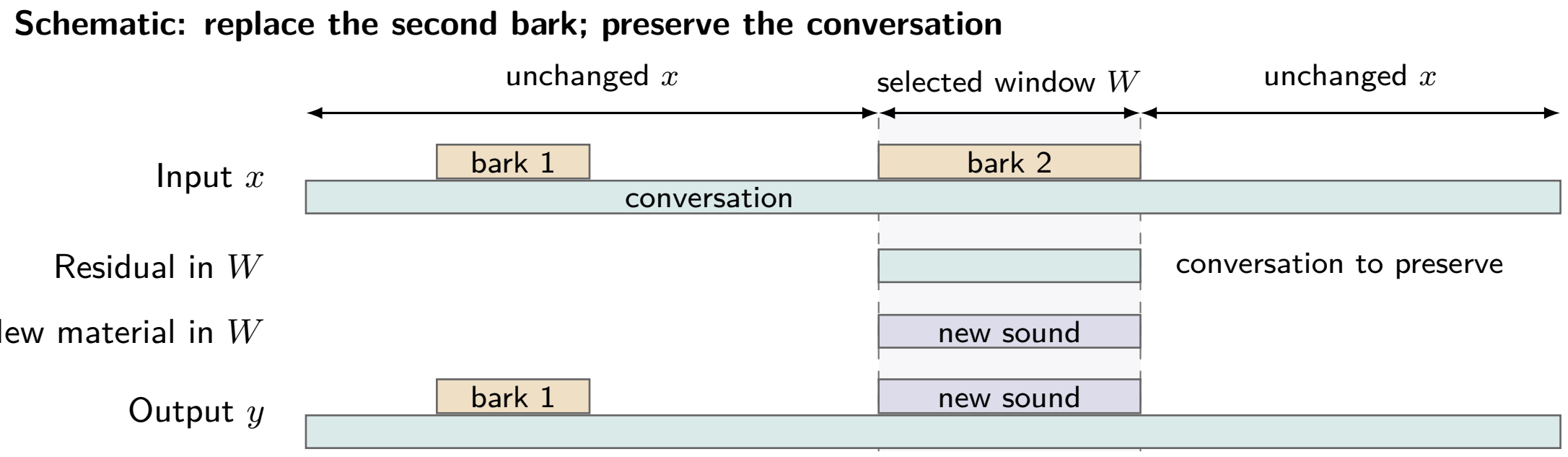

\subsection{Evidence-guided execution}
\label{sec:evidence_execution}

Correct planning alone does not ensure that intermediate audio supports the intended edit. Replacement, for example, depends on both removing the old sound and obtaining suitable new material. Harmony therefore conditions corrective actions on the instruction, the plan, and observations of the first separation or generation outputs.

Let $x$ denote the original recording and $\pi$ the edit plan produced during planning. For corrective action $j$, $z_j$ is the intermediate audio before correction, and $e_j^{(0)}$ contains observations of the original audio and the initial separation or generation result. The resulting audio is
\begin{equation}
\tilde z_j =
\begin{cases}
T_j\!\left(x,z_j;\pi,e_j^{(0)}\right),
& c_j=1, \\
z_j, & c_j=0.
\end{cases}
\label{eq:conditional_execution}
\end{equation}
Here, $T_j$ is the corrective tool operation. The binary decision $c_j$ equals 1 when the action is applicable and the observations support its use under the plan's editing and preservation requirements. Applicable actions can be composed in execution order.

Harmony uses audio evidence to select among predefined corrective actions on separated or generated material. These actions address specific mismatches while retaining the planned target, execution windows, and preservation requirements. Table~\ref{tab:correction_scope} summarizes the evidence supporting each action.

Corrections follow fixed, action-specific limits: separation-query correction permits one additional separation for supported whole-recording knocking or noise removal, while generated-asset cleanup allows one correction of each type per asset. For timbre correction, Basic Pitch~\citep{basicpitch} extracts the generated asset's note sequence and timing for a single rendering with the target timbre, retaining its transcribed musical structure.

\subsection{Temporal composition and mixing}
\label{sec:execution}

After any applicable correction, Harmony assembles the output within the execution windows $W$ (Figure~\ref{fig:mixing}). Addition overlays new material on the original recording; removal uses the separated residual; replacement combines the residual with new material. Volume adjustment applies the requested gain to the selected source or mixture. For edits that retain the original timeline, these changes are confined to $W$, with smooth transitions at window boundaries; audio outside the windows is copied from $x$.

Placement uses each asset's actual duration to apply the requested position, count, spacing, and coverage constraints. Replacement shares the same windows between removal and insertion, aligning the new material with the selected occurrences. Duration changes or shifts of subsequent content are applied when requested by the instruction.

For additive mixing, the original track retains unit gain. Eligible continuous context layers at default levels receive local peak control: only the added layer is attenuated where the mixture needs additional headroom. Explicit gain settings follow the instruction.

\section{Experiments}
\label{sec:experiments}

\begingroup
\setlength{\textfloatsep}{10pt plus 2pt minus 2pt}
\setlength{\floatsep}{10pt plus 2pt minus 2pt}
\setlength{\intextsep}{10pt plus 2pt minus 2pt}
\makeatletter
\setlength{\@fptop}{0pt}
\setlength{\@fpsep}{12pt}
\setlength{\@fpbot}{0pt plus 1fil}
\makeatother

\paragraph{Dataset and setup.}
We evaluate all 420 MMAE examples~\citep{ma2026mmae} that fall into one of four operation categories, including addition, removal, replacement, and volume adjustment. The set contains 115 addition, 149 removal, 79 replacement, and 77 volume-adjustment examples; input recordings may contain speech. The evaluation set contains 176 recordings of at most 10 seconds and 244 recordings longer than 10 seconds. Harmony uses \texttt{gpt-5-mini-2025-08-07} as its LLM planner with reasoning effort set to \texttt{low}. Long recordings are provided in full, without external segmentation into shorter clips. We used MMAE examples outside the 420-example evaluation set for development and debugging.

\paragraph{Implementation.}
We use Qwen3-Omni-30B-A3B-Instruct for audio observation, SAM Audio large~\citep{samaudio} for separation and generated-asset cleanup, Whisper large-v3-turbo~\citep{whisper} for transcription, and Stable Audio 3~\citep{stableaudio3} for generation. SAM Audio retains its internal eight-candidate reranking and predicts separation spans when explicit anchors are absent. Music rendering uses Basic Pitch notes~\citep{basicpitch} or, for supported continuous-pitch transformations, pYIN contours~\citep{pyin}, with the GeneralUser GS SoundFont.\footnote{\url{https://www.schristiancollins.com/generaluser.php}}

Planner requests do not explicitly set temperature, top-$p$, or seed. Each pre-edit controller invocation allows up to two audio observations and two analysis-tool calls, using each tool type at most once and performing at most one separation. Beyond this initial budget, recovery permits one additional planning attempt per addition context, one additional grounding attempt per removal plan, and one grounding attempt per unresolved volume action. Execution corrections follow the limits in Section~\ref{sec:evidence_execution}.

When addition timing remains unresolved, Harmony may use an explicitly marked estimate while preserving the original timeline and base gain. User-specified constraints remain binding, and replacement uses its established execution windows.

\paragraph{Baselines.}
We compare against SmartDJ~\citep{lan2025smartdj}, MMEDIT~\citep{tao2025mmedit}, WavCraft~\citep{liang2024wavcraft}, and the Audio Editing Challenge Agent-track baseline (AEC Agent)~\citep{aec2026baseline}. For WavCraft, we report both its original configuration and a matched-backbone configuration in which the LLM and all branch models are replaced with their Harmony counterparts. This configuration supports a comparison using the same underlying model choices. AEC Agent uses its original DeepSeek-V4-Flash router with DSP tools, SAM Audio separation, and AuK generative editing with Prompt Enhancer enabled. We retain its released prompts, tools, and parameters.

\paragraph{Evaluation protocol.}
We use MMAE's rubric-based evaluator~\citep{ma2026mmae}, with Qwen3-Omni as an external judge. Each rubric is a multiple-choice question about a requested change or preservation requirement, using the input and/or edited audio specified by that criterion. We evaluate each rubric three times with shuffled option order. It receives a binary pass when at least two responses match the annotated correct option.

\paragraph{Metrics.}
We report three complementary metrics, all as percentages:
\begin{enumerate}
\setlength{\itemsep}{3pt}
\setlength{\parsep}{0pt}
\item \emph{Instruction Following Rate (IFR)} measures fulfillment of the requested changes. For each example, we compute the fraction of instruction-following rubrics passed, then average these fractions across examples. Higher IFR indicates that more of the requested edit requirements are satisfied on average.
\item \emph{Consistency Rate (CR)} measures preservation of content that should remain unchanged. It is computed in the same way using the consistency rubrics. Higher CR indicates better fulfillment of preservation requirements, such as retaining background sounds during a targeted removal.
\item \emph{Exact Match Rate (EMR)} is the percentage of examples passing every instruction-following and consistency rubric. It measures complete success under the evaluator: one failed criterion is sufficient to exclude an example. Here, ``exact match'' means that all criteria pass, not that two waveforms are identical.
\end{enumerate}

For example, an output that preserves the background but leaves the target laughter in a replacement edit may score well on CR while failing EMR. Reading the three metrics together distinguishes partial fulfillment from complete edits.


\begin{table}[t]
\centering
\setlength{\tabcolsep}{5pt}
\renewcommand{\arraystretch}{1.02}
\caption{Results by duration and overall (\%, $\uparrow$). Overall scores include task-macro and pooled averages; duration groups use pooled averages. Bold marks the best score in each column.}
\label{tab:results_duration}
\begin{tabular}{@{}lrrrrrr@{}}
\toprule
& \multicolumn{3}{c}{At most 10 s ($n=176$)} & \multicolumn{3}{c}{Longer than 10 s ($n=244$)} \\
\cmidrule(lr){2-4}\cmidrule(l){5-7}
Method & IFR & CR & EMR & IFR & CR & EMR \\
\midrule
SmartDJ & 50.74 & 65.58 & 7.39 & 54.70 & 63.84 & 5.33 \\
MMEDIT & 53.35 & 66.57 & 6.25 & 19.97 & \textbf{98.24} & 2.05 \\
WavCraft (original) & 59.23 & 81.38 & 17.05 & 57.65 & 75.75 & 11.07 \\
WavCraft (matched backbone) & 54.96 & 83.28 & 12.50 & 50.15 & 85.25 & 12.30 \\
AEC Agent & 44.68 & 75.60 & 7.95 & 44.13 & 74.12 & 8.61 \\
Harmony (ours) & \textbf{60.71} & \textbf{89.71} & \textbf{23.86} & \textbf{63.24} & 86.27 & \textbf{21.72} \\
\midrule
& \multicolumn{3}{c}{Overall: task macro} & \multicolumn{3}{c}{Overall: pooled ($n=420$)} \\
\cmidrule(lr){2-4}\cmidrule(l){5-7}
Method & IFR & CR & EMR & IFR & CR & EMR \\
\midrule
SmartDJ & 52.34 & 65.82 & 5.68 & 53.04 & 64.57 & 6.19 \\
MMEDIT & 33.93 & 83.29 & 3.64 & 33.96 & 84.97 & 3.81 \\
WavCraft (original) & 56.75 & 77.65 & 12.76 & 58.31 & 78.11 & 13.57 \\
WavCraft (matched backbone) & 51.05 & 83.42 & 11.52 & 52.17 & 84.43 & 12.38 \\
AEC Agent & 42.98 & 72.23 & 6.43 & 44.36 & 74.74 & 8.33 \\
Harmony (ours) & \textbf{61.16} & \textbf{87.21} & \textbf{21.50} & \textbf{62.18} & \textbf{87.71} & \textbf{22.62} \\
\bottomrule
\end{tabular}
\end{table}

\begin{table}[t]
\centering
\setlength{\tabcolsep}{5pt}
\renewcommand{\arraystretch}{1.02}
\caption{Results by editing operation. All scores are percentages ($\uparrow$); bold indicates the best score within each operation.}
\label{tab:results_operation}
\begin{tabular}{@{}lrrrrrr@{}}
\toprule
& \multicolumn{3}{c}{Addition ($n=115$)} & \multicolumn{3}{c}{Removal ($n=149$)} \\
\cmidrule(lr){2-4}\cmidrule(l){5-7}
Method & IFR & CR & EMR & IFR & CR & EMR \\
\midrule
SmartDJ & 57.42 & 69.67 & 9.57 & 53.55 & 56.95 & 6.71 \\
MMEDIT & 39.13 & 82.56 & 6.09 & 31.36 & \textbf{93.94} & 3.36 \\
WavCraft (original) & 70.25 & 78.31 & 21.74 & \textbf{58.19} & 80.58 & 12.75 \\
WavCraft (matched backbone) & 62.80 & \textbf{83.70} & 17.39 & 51.01 & 89.45 & 13.42 \\
AEC Agent & 30.68 & 71.01 & 4.35 & 57.60 & 87.75 & 18.79 \\
Harmony (ours) & \textbf{79.14} & 82.00 & \textbf{31.30} & 57.29 & 93.12 & \textbf{22.82} \\
\midrule
& \multicolumn{3}{c}{Replacement ($n=79$)} & \multicolumn{3}{c}{Volume adjustment ($n=77$)} \\
\cmidrule(lr){2-4}\cmidrule(l){5-7}
Method & IFR & CR & EMR & IFR & CR & EMR \\
\midrule
SmartDJ & 58.92 & 50.22 & 2.53 & 39.46 & 86.44 & 3.90 \\
MMEDIT & 34.98 & 66.55 & 2.53 & 30.24 & 90.10 & 2.60 \\
WavCraft (original) & \textbf{77.13} & 55.63 & 12.66 & 21.43 & \textbf{96.09} & 3.90 \\
WavCraft (matched backbone) & 64.66 & 71.46 & 12.66 & 25.71 & 89.08 & 2.60 \\
AEC Agent & 39.05 & 63.92 & 1.27 & \textbf{44.61} & 66.26 & 1.30 \\
Harmony (ours) & 72.24 & \textbf{78.34} & \textbf{22.78} & 35.97 & 95.40 & \textbf{9.09} \\
\bottomrule
\end{tabular}
\end{table}

\paragraph{Main results.}
Table~\ref{tab:results_duration} reports scores by input duration and overall; Table~\ref{tab:results_operation} reports each operation separately. Pooled averages give equal weight to examples, whereas task-macro averages give equal weight to the four operations. All examples are included. Duration groups are determined from original input frame counts and reuse the same evaluation scores.

Harmony achieves the highest pooled IFR (62.18\%), CR (87.71\%), and EMR (22.62\%) among the compared methods (Table~\ref{tab:results_duration}). Its EMR corresponds to 95 of 420 examples passing every editing and preservation criterion, compared with 57 for original WavCraft. The improvement therefore extends to completing the requested edit as a whole, alongside higher average fulfillment of editing and preservation requirements. Harmony also leads in IFR and EMR in both duration groups, indicating that this advantage is present for both short and longer recordings.

High IFR or CR alone does not establish complete editing success: an output may fulfill the requested change while altering content that should remain, or preserve the recording without completing the edit. EMR addresses both requirements by counting an example as successful only when all editing and preservation criteria pass. Different editing operations also place different demands on the component models and their coordination. Harmony achieves the highest EMR across all four operations (Table~\ref{tab:results_operation}), indicating that its advantage in complete editing success extends across these different demands.


\begingroup
\newcommand{\harmonylisten}[2]{%
  \href{https://christong9.github.io/harmony/\#case-#1}{%
    \tikz[baseline=(audiolink.base)]{%
      \node[draw=blue!40!black,fill=blue!4,rounded corners=2pt,
        inner xsep=7pt,inner ysep=4pt,text=blue!50!black,
        font=\sffamily\small\bfseries] (audiolink)
        {$\blacktriangleright$\ #2};%
    }%
  }%
}
\begin{figure}[!ht]
\centering
\includegraphics[width=\linewidth]{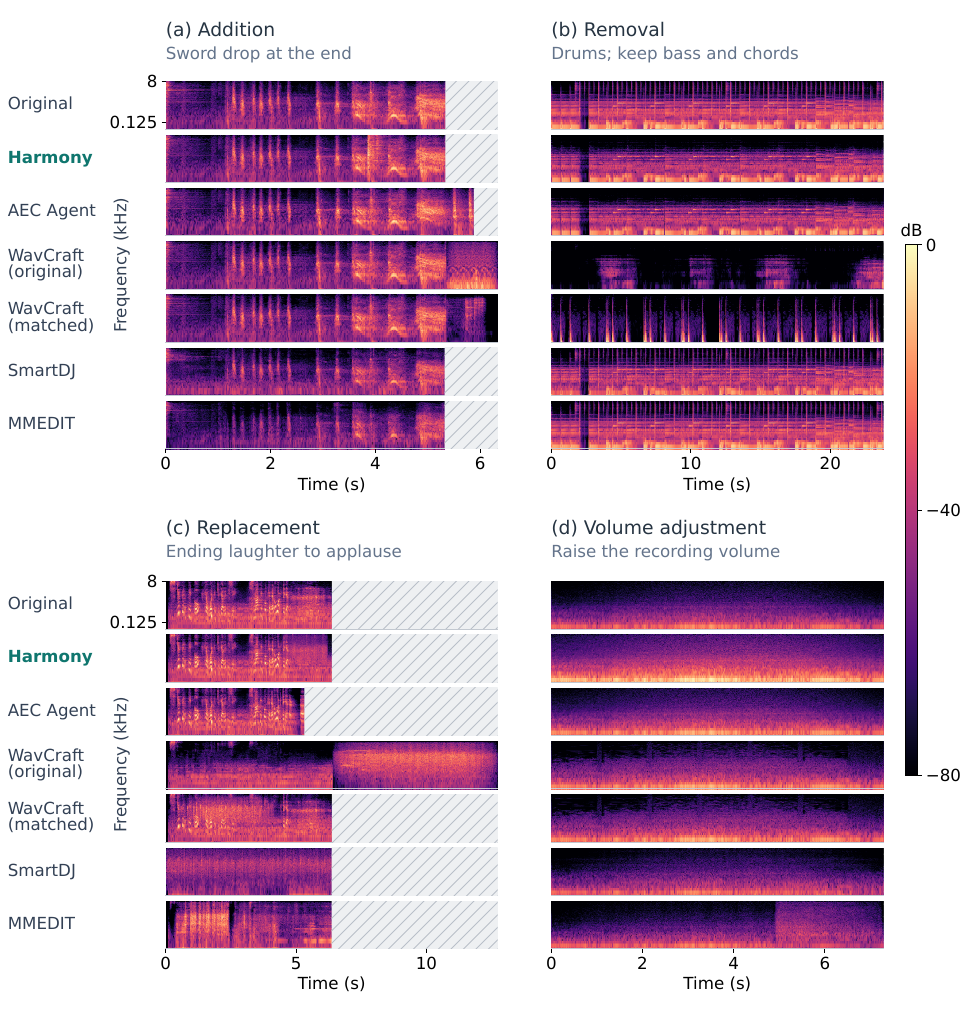}\par
\smallskip
{\small\bfseries Listen:}\hspace{0.5em}%
\harmonylisten{69e8980c910050e59545dd0f}{(a) Addition}\hspace{0.6em}%
\harmonylisten{69e89805bd490e32ff0d3c8e}{(b) Removal}\hspace{0.6em}%
\harmonylisten{69e89817910050e595461416}{(c) Replacement}\hspace{0.6em}%
\harmonylisten{69e8981f6508941e55c185f1}{(d) Volume}
\caption{Four editing operations. Each case shares its color reference across methods. Hatching marks time beyond a file's end.}
\label{fig:qualitative_spectrograms}
\end{figure}

\begin{figure}[!ht]
\centering
\includegraphics[width=\linewidth]{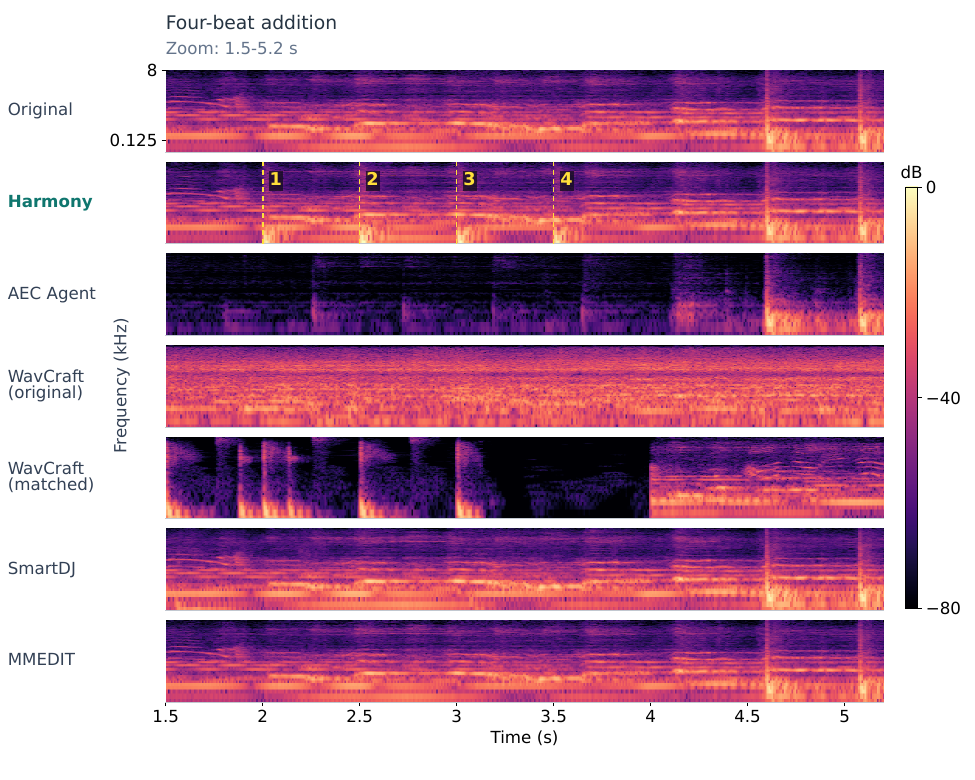}\par
\smallskip
\harmonylisten{69e89805bd490e32ff0d3c9d}{Listen: four drum beats}
\caption{Four-beat addition, enlarged over 1.5--5.2 s. Numbered markers indicate the four additions; all methods share one color reference from their full recordings.}
\label{fig:qualitative_four_beats}
\end{figure}

\subsection{Qualitative analysis}
\label{sec:qualitative}

Figure~\ref{fig:qualitative_spectrograms} illustrates the changes and preservation requirements underlying IFR and CR across the four operations. Addition (a) and replacement (c) modify the ending while preserving the first 3.8 and 4.4 seconds, respectively, sample for sample. Removal (b) attenuates transient structures while retaining sustained bands. Volume adjustment (d) retains the time--frequency structure, with waveform fitting confirming an approximately uniform $+15$ dB gain. The linked audio complements these observations by allowing comparison of sound identity and residual content.

In the ending-laughter replacement example in Figure~\ref{fig:qualitative_spectrograms}(c), SmartDJ turns the entire output into applause, removing content beyond the requested ending. Harmony replaces the ending laughter while keeping the preceding audio intact, allowing the requested change to fit into the existing recording.

Figure~\ref{fig:qualitative_four_beats} tests an explicit event-count requirement. Harmony adds four transients at 2.0, 2.5, 3.0, and 3.5 seconds, while samples outside the four addition windows remain unchanged. Matched-backbone WavCraft does not produce the requested number of beats. Listening also reveals lower-quality drum sounds in original WavCraft, highlighting the importance of producing suitable audio as well as arranging it according to the instruction. This example illustrates how event placement and context preservation can be inspected together.

Additional examples are available on the \href{https://christong9.github.io/harmony/}{\underline{project demo website}}.
\endgroup

\begin{table}[t]
\centering
\setlength{\tabcolsep}{5pt}

\renewcommand{\arraystretch}{1.02}
\caption{Task-level $2\times2$ ablation. Bold indicates the highest metric within each task.}
\label{tab:ablation_operation}
\begin{tabular}{@{}lrrrrrr@{}}
\toprule
& \multicolumn{3}{c}{Addition ($n=115$)} & \multicolumn{3}{c}{Removal ($n=149$)} \\
\cmidrule(lr){2-4}\cmidrule(l){5-7}
Configuration & IFR & CR & EMR & IFR & CR & EMR \\
\midrule
$A_0B_0$ & 77.09 & \textbf{83.68} & 28.70 & 55.53 & 92.53 & 21.48 \\
$A_1B_0$ & 79.10 & 81.52 & 27.83 & 60.95 & 90.86 & 22.15 \\
$A_0B_1$ & 73.28 & 81.51 & 18.26 & \textbf{61.73} & \textbf{94.00} & \textbf{26.17} \\
$A_1B_1$ & \textbf{79.14} & 82.00 & \textbf{31.30} & 57.29 & 93.12 & 22.82 \\
\midrule
& \multicolumn{3}{c}{Replacement ($n=79$)} & \multicolumn{3}{c}{Volume adjustment ($n=77$)} \\
\cmidrule(lr){2-4}\cmidrule(l){5-7}
Configuration & IFR & CR & EMR & IFR & CR & EMR \\
\midrule
$A_0B_0$ & 65.27 & \textbf{80.84} & 16.46 & 30.89 & \textbf{95.85} & 3.90 \\
$A_1B_0$ & 68.80 & 80.55 & 20.25 & 32.32 & 95.64 & 6.49 \\
$A_0B_1$ & 69.45 & 75.21 & 12.66 & 29.05 & 95.11 & 5.19 \\
$A_1B_1$ & \textbf{72.24} & 78.34 & \textbf{22.78} & \textbf{35.97} & 95.40 & \textbf{9.09} \\
\bottomrule
\end{tabular}
\end{table}

\begin{table}[t]
\centering
\setlength{\tabcolsep}{6pt}

\renewcommand{\arraystretch}{1.02}
\caption{Aggregate ablation results. Bold indicates the highest score in each column.}
\label{tab:ablation_average}
\begin{tabular}{@{}lrrrrrr@{}}
\toprule
& \multicolumn{3}{c}{Macro} & \multicolumn{3}{c}{Pooled} \\
\cmidrule(lr){2-4}\cmidrule(l){5-7}
Configuration & IFR & CR & EMR & IFR & CR & EMR \\
\midrule
$A_0B_0$ & 57.19 & \textbf{88.23} & 17.63 & 58.75 & \textbf{88.52} & 19.29 \\
$A_1B_0$ & 60.29 & 87.14 & 19.18 & 62.15 & 87.24 & 20.48 \\
$A_0B_1$ & 58.38 & 86.46 & 15.57 & 60.35 & 87.25 & 17.62 \\
$A_1B_1$ & \textbf{61.16} & 87.21 & \textbf{21.50} & \textbf{62.18} & 87.71 & \textbf{22.62} \\
\bottomrule
\end{tabular}
\end{table}

\subsection{Ablation study}
\label{sec:ablation}

We independently enable or disable external evidence acquisition ($A$) and conditional correction ($B$) in a $2\times2$ ablation. $A$ makes external activity, boundary, source-activity, and transcription tools available during planning; enabling it does not require a tool call for every input. Qwen-based listening and LLM planning remain active when $A$ is disabled. All configurations retain the underlying separation, generation, and mixing workflow and are evaluated on the same complete sample set for each operation.

External evidence improves instruction fulfillment across all four operations when correction is disabled. This supports using external analysis tools alongside Qwen's observations to inform event selection and temporal planning. However, the improvement does not always translate into complete editing success: addition achieves higher IFR without higher EMR. Satisfying more individual instruction criteria is therefore useful, but the resulting edit still needs to meet the remaining editing and preservation requirements.

Correction provides a complementary benefit when external evidence is available. Enabling it improves EMR in every operation, while pooled IFR remains almost unchanged. Its contribution is thus more apparent in the number of edits that satisfy all requirements together than in average instruction fulfillment. Without external tools, however, correction lowers pooled EMR, particularly for addition and replacement. This contrast suggests that additional corrective actions alone are insufficient; their effectiveness depends on the information supporting the editing decisions. The results support coordinating evidence acquisition and correction throughout the workflow.

The joint configuration achieves the highest aggregate IFR and EMR (Table~\ref{tab:ablation_average}), with the strongest complete-success results for addition, replacement, and volume adjustment (Table~\ref{tab:ablation_operation}). Its slightly lower CR than the configuration with both mechanisms disabled shows that this improvement is accompanied by a preservation trade-off under the evaluation criteria. Nevertheless, higher EMR means that more examples meet both editing and preservation requirements in full. Removal is an exception, performing best with correction alone, so the benefit of combining the mechanisms varies across operations.

\begin{table}[!htbp]
\centering
\setlength{\tabcolsep}{12pt}
\caption{Mean cost per input in the first full executions. Overall values weight all 420 inputs equally. Subsequent offline repairs are excluded.}
\label{tab:runtime}
\begin{tabular}{@{}lrrr@{}}
\toprule
Operation & $n$ & Time (s) & LLM calls \\
\midrule
Addition & 115 & 52.35 & 3.71 \\
Removal & 149 & 115.36 & 3.69 \\
Replacement & 79 & 173.08 & 7.04 \\
Volume adjustment & 77 & 95.36 & 3.29 \\
\midrule
Overall & 420 & 105.30 & 4.25 \\
\bottomrule
\end{tabular}
\end{table}
\subsection{Computational cost}
\label{sec:cost}

Table~\ref{tab:runtime} reports the first full executions of all 420 inputs, including unsuccessful attempts. Timing includes input handling, editing, and output writing, but excludes evaluation, queueing, service startup, and initialization outside the timed routine. LLM counts cover logged routing, planning, and review requests. Replacement requires the most time and calls. Hardware and service load were not standardized for speed comparisons.

\subsection{User study}
\label{sec:user_study}Sixteen participants compared anonymous outputs from Harmony, WavCraft (original), and AEC Agent on 16 curated examples, providing 128 judgments of overall editing (correctness and listening quality) and 120 applicable preservation judgments. Participants answered two questions: (1) ``Which result best fulfills the editing request overall, considering both correctness and listening quality?'' and (2) ``Which result best preserves the sounds and time regions that the instruction did not ask to change?'' Harmony received the most preferences on both criteria (Table~\ref{tab:user_study}).


\begin{table}[!ht]
\centering
\setlength{\tabcolsep}{12pt}
\caption{Listener preferences (\%). Ties and cannot-judge responses are included; inapplicable responses are excluded.}
\label{tab:user_study}
\begin{tabular}{@{}lcc@{}}
\toprule
Method & Overall editing ($n=128$) & Preservation ($n=120$) \\
\midrule
WavCraft (original) & 9.38 & 9.17 \\
AEC Agent & 8.59 & 16.67 \\
Harmony (ours) & \textbf{69.53} & \textbf{56.67} \\
\bottomrule
\end{tabular}
\end{table}

\endgroup

\section{Limitations and Future Work}
\label{sec:limitations}

Harmony often needs several LLM calls to plan complex edits, increasing runtime and cost. Even when planning succeeds, the edited audio may not blend naturally with the original recording, leading to abrupt transitions. In terms of task scope, the current system supports single-step requests but cannot yet handle instructions that require several edits in sequence. More broadly, editing requests can involve many different sounds, timing constraints, and combinations of operations, so the generalization ability of our model could be further improved.

Harmony currently operates on audio alone. Future work will explore joint audio--video editing through coordinated perception, analysis, and editing tools, using audiovisual evidence to identify corresponding events and maintain temporal consistency.

We will also investigate reinforcement learning for evidence acquisition and correction. The current controller follows fixed rules and may use an explicitly marked estimate when addition timing remains unresolved. Learning tool selection and correction decisions from instruction-fulfillment and preservation feedback could make this process more adaptive.

\section{Conclusion}
We presented Harmony, an agentic framework that bridges audio understanding and editing through coordinated planning and execution. Harmony introduces two complementary mechanisms: edit-driven collaborative planning, which combines audio-language understanding with external analysis tools to identify editing targets, temporal boundaries, and preservation requirements; and evidence-guided execution, which uses observations of intermediate audio to guide bounded corrections before final mixing. Together, these mechanisms enable more precise control over what to modify and what to preserve. Experiments on 420 MMAE examples across four editing operations demonstrate that Harmony outperforms existing approaches in overall instruction following, context preservation, and exact match rate, improving exact match rate from 13.57\% to 22.62\% over the strongest baseline. Ablation studies further highlight the complementary roles of evidence acquisition and conditional correction. Our findings demonstrate the value of integrating audio understanding with editing operations and provide a foundation for developing more reliable and flexible natural-language audio editing systems.

\section*{Broader Impact}
Audio editing can support creativity but also misrepresent events. Users should respect source rights, privacy, and consent, protect private speech, use licensed or consented demonstration material, and disclose edits when authenticity matters.

\bibliography{references_related,references_models}
\bibliographystyle{tmlr}
\end{document}